%% file: Gadolinio.tex
\documentclass[a4paper,11pt, noamssymbols]{article}
\usepackage[utf8]{inputenc}

\pdfoutput=1 
\usepackage{jinstpub} 
\usepackage{graphicx}
\usepackage{multirow}
\usepackage{cancel}
\usepackage{soul}
\usepackage{subcaption}
\usepackage{fancyhdr}
\usepackage[version=4]{mhchem}
\usepackage{gensymb}
\usepackage{xr}
\usepackage{longtable}
\usepackage{color}
\usepackage{url}
\usepackage{textcomp}
\usepackage{appendix}
\usepackage{bm}
\usepackage{comment}
\usepackage{hhline}
\usepackage{lineno}
\makeatletter
\usepackage{scrextend}
\usepackage{tablefootnote}

\title{\boldmath A new hybrid gadolinium nanoparticles-loaded polymeric material for neutron detection in rare event searches}

\emailAdd{bianca.bottino@ge.infn.it}
\emailAdd{anna.marini@ge.infn.it}

\abstract{
Experiments aimed at direct searches for WIMP dark matter require highly effective
 reduction of backgrounds and control of any residual radioactive 
contamination. In particular, neutrons interacting with atomic nuclei represent
 an important class of backgrounds due to the expected similarity of a WIMP-nucleon 
interaction, so that such experiments often feature a dedicated neutron detector 
surrounding the active target volume. In the context of the development of DarkSide-20k detector at INFN Gran Sasso National Laboratory (LNGS), several R\&D projects were conceived and developed for the creation of a new hybrid material rich in both hydrogen and gadolinium nuclei to be employed as an essential element of the neutron detector. 
Thanks to its very high cross-section for neutron capture, gadolinium is one of the most widely used elements in neutron detectors, while the hydrogen-rich material is instrumental in efficiently moderating the neutrons. In this paper results from one of the R\&Ds are presented. 
In this effort the new hybrid material was obtained as a poly(methyl methacrylate) (PMMA) matrix, loaded with gadolinium oxide in the form of nanoparticles. We describe its realization, including all phases of design, purification, construction, characterization, and determination of mechanical properties of the new material.}

\keywords{Only keywords from JINST's keywords list please}

\arxivnumber{XXXXX} 

\begin{document}

\input{autori}
\maketitle

\flushbottom
\input{Motivation_new}

\input{Requirements_new}

\input{Radiopurity_new}

\input{Experimental_part_new}

\input{Characterization_new}
\input{Industrial_tests_new}
\input{Conclusions}

\acknowledgments
The Authors express their gratitude to G. Sobrero, F. Soggia, R. Fiorini, Prof. P. Manfrinetti, Prof. T. Benelli and Prof. L. Mazzocchetti for the constant techincal support to the project. The Authors would also like to thank the mechanical workshop of the INFN Genova section, in particular P. Pollovio, and the technicians from LNGS and LSC. \\
This work has been supported by São Paulo's Research Foundation (FAPESP) Grant 2021/11489-7; Ivone Albuquerque and Edivaldo M. Santos are partially supported by the Brazilian CNPq. B. Costa and R. Perez are supported by  FAPESP and L. Kerr by CNPq.\\
Support is acknowledged by the Deutsche Forschungsgemeinschaft (DFG, German Research Foundation) under Germany’s Excellence Strategy – EXC 2121 „Quantum Universe“ – 390833306.\\
We acknowledge the financial support by the Chinese Academy of Sciences (113111KYSB20210030) and National Natural Science Foundation of China (12020101004). This work was supported by the Spanish Ministerio de Ciencia, Innovación y Universidades  with the grant PID2022-138357NB-C22.



\appendix

\include{Appendix}\label{Appendix}

\end{document}

%% file: autori.tex
\author[1] {F.~Acerbi}
\affiliation[1]{Fondazione Bruno Kessler, Povo 38123, Italy}
\author[2]{P.~Adhikari}
\affiliation[2]{Department of Physics, Carleton University, Ottawa, ON K1S 5B6, Canada}
\author[3,4]{P.~Agnes}
\affiliation[3]{Gran Sasso Science Institute, L'Aquila 67100, Italy}
\affiliation[4]{INFN Laboratori Nazionali del Gran Sasso, Assergi (AQ) 67100, Italy}
\author[5] {I.~Ahmad}
\affiliation[5]{AstroCeNT, Nicolaus Copernicus Astronomical Center of the Polish Academy of Sciences, 00-614 Warsaw, Poland}
\author[6,7]{S.~Albergo}
\affiliation[6]{INFN Catania, Catania 95121, Italy}
\affiliation[7]{Universit\`a of Catania, Catania 95124, Italy}
\author[8]{I.~F.~Albuquerque}
\affiliation[8]{Instituto de F\'isica, Universidade de S\~ao Paulo, S\~ao Paulo 05508-090, Brazil}
\author[9]{T.~Alexander}
\affiliation[9]{Pacific Northwest National Laboratory, Richland, WA 99352, USA}
\author[10]{A.~K.~Alton}
\affiliation[10]{Physics Department, Augustana University, Sioux Falls, SD 57197, USA}
\author[11]{P.~Amaudruz}
\affiliation[11]{TRIUMF, 4004 Wesbrook Mall, Vancouver, BC V6T 2A3, Canada}
\author[3,4]{M.~Angiolilli}
\author[12]{E.~Aprile}
\affiliation[12]{Physics Department, Columbia University, New York, NY 10027, USA}
% ...

%% file: Motivation_new.tex
\section{Introduction}

\label {sec:Introduction} 
This paper describes one of the parallel R\&D projects carried out by the DarkSide collaboration for the development of a new neutron-tagging material made of poly(methyl methacrylate),
PMMA $(C_5 O_2 H_8)_n$, loaded with up to a few percent of gadolinium (\ce{Gd}) by weight, referred to as \ce{Gd}-PMMA in the following.
This work has been driven by the requirements (Sec.~\ref{sec:requirements}) of DarkSide-20k (for a description of the detector see ~\cite{DSRefCosmogenicActivation}, an experiment aiming at the direct detection of WIMPs (Weakly Interacting Massive Particles) dark matter~\cite{Bertone} at the Gran Sasso National Laboratory (LNGS), in Italy. Since the low background level is one of the most important features for the experiment, as a first step we carried out extensive market research and screening of the radioactivity of the materials used in this R\&D, in order to select the most suitable (Sec.~\ref{sec:Radiopurity}). We have developed a method of production of \ce{Gd}-PMMA with an extensive series of laboratory tests (Sec.~\ref{section:LabTests}) and we also set up and optimized a set of characterization measurements of the  samples produced (Sec.~\ref{sec:char}). When the results obtained on laboratory samples were satisfactory, the process was transferred and adapted to industrial production 
(Sec.~\ref{sec:IndustrialTests}). The same procedures as developed during the laboratory test phase were used to perform the quality assurance of the industrially produced sheets.\\
 
The following considerations informed the development of the \ce{Gd}-PMMA:
\begin{itemize}
\item the need for integration into the DarkSide-20k detector at LNGS, including operation in liquid argon at 87 K and full containment of \ce{Gd} ensuring it can not be released into the environment;
\item the need for a homogeneous distribution of gadolinium, of sufficient concentration, within the \ce{Gd}-PMMA to ensure the neutron tagging is efficient throughout the detector; 
\item the need for minimal radioactivity of the product to keep the DarkSide-20k background within design limits, affecting the choice of primary ingredients and production procedures; 
\item the need to produce large quantities of \ce{Gd}-PMMA, of the order of 20 t, using a \ce{Gd} compound easily available on the market.
\end{itemize}
When appropriately coupled to standard commercial gamma-ray detectors,
this material may be suitable for general purpose neutron detectors.
Given its scalability and relatively low cost, our \ce{Gd}-PMMA has excellent features for a neutron tagging material, not only for DarkSide-20k but also for other large detectors searching for rare events.

%% file: Requirements_new.tex
\section{Requirements}
\label{sec:requirements}
\subsection{Integration into DarkSide-20k}

\begin{figure}[!h]
    \centering
    \includegraphics[width=0.65\textwidth]{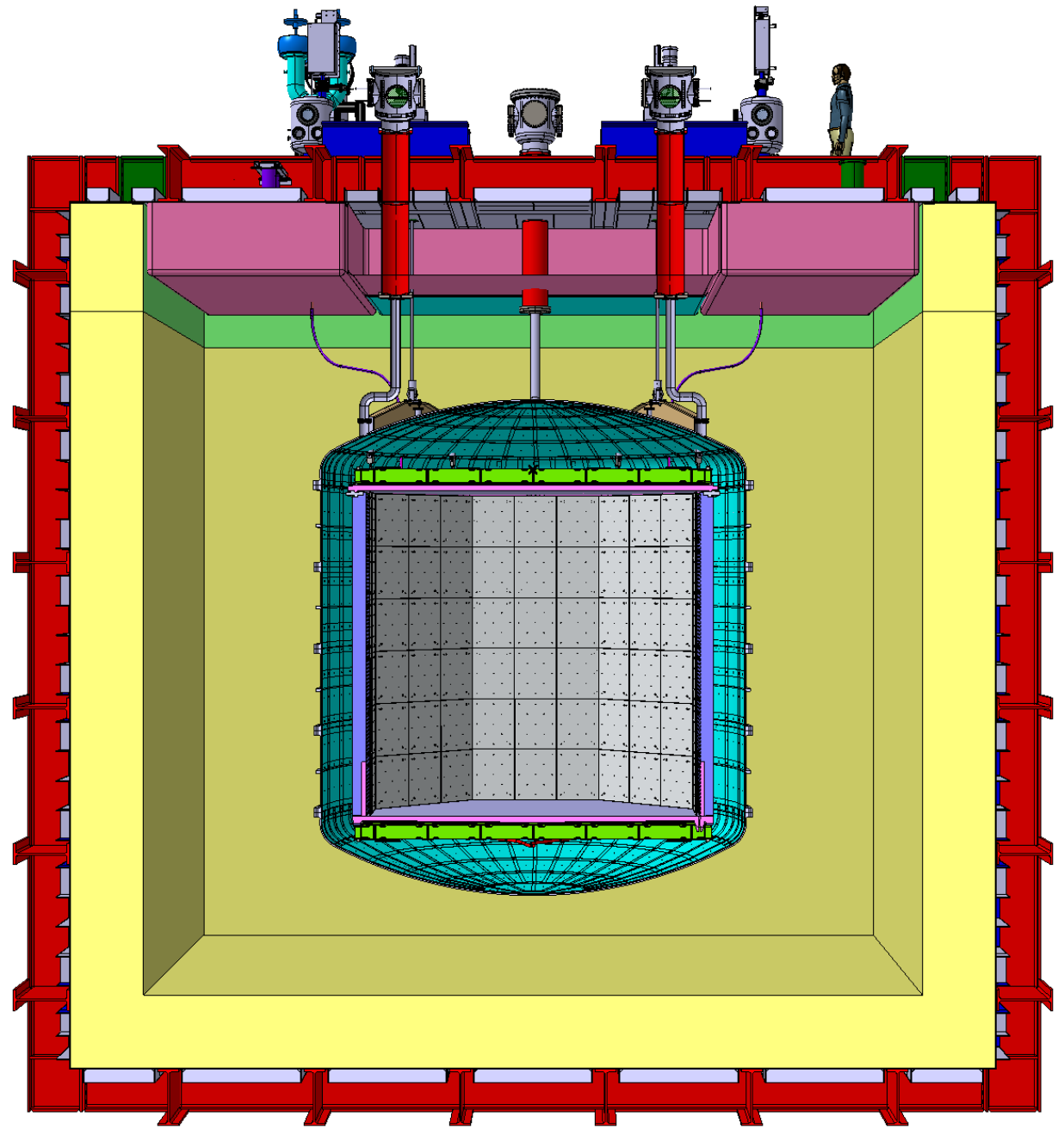}
    \caption{The picture shows a drawing of the DarkSide-20k apparatus. The innermost part is the TPC: the position of the Gd-PMMA lateral walls (which form an octagon) is shown in violet, while the pure PMMA windows at the top and bottom are depicted in pink. Above and below the pure PMMA windows there are the Gd-PMMA endcaps, which are represented in bright green. The TPC and Veto are contained in a stainless steel vessel, light blue in the picture, that holds the underground liquid argon. The vessel is hanging from the top of a big membrane cryostat (in red and yellow), that will be filled with atmospheric liquid argon. Neutrons moderated and captured in the Gd-PMMA (violet and green), produce $\gamma$ rays that induce scintillation in the surrounding liquid argon, therefore they are tagged.}
    \label{fig:DS_sketch}
 \end{figure}

The center of the DarkSide-20k detector is a dual-phase liquid argon Time Projection Chamber (TPC) instrumented to detect scintillation photons produced in primary excitation of the argon and from electroluminescence produced by the ionization electrons in the gas region at the top of the TPC. The TPC is submerged in 100 t of low-radioactivity argon, extracted from underground sources. As demonstrated by the predecessor experiment DarkSide-50, in underground argon the level of the $\beta$-radioactive isotope \ce{^{39}Ar} is lower by more than a factor 1400 than in the standard argon of atmospheric origin \cite{ds50_1,ds50_u, ds50_500}. 
The goal of the experiment is to observe WIMPs scattering elastically off the argon nucleus, whose recoil deposits tens to hundreds of keV of energy in the material. Fig.~\ref{fig:DS_sketch} shows a sketch of the DarkSide-20k detector with some details.
The detector is designed for a 200 t$\cdot$yr exposure with a negligible
instrumental background level in the WIMP search region of interest.
While argon has excellent rejection capabilities against electromagnetic backgrounds, neutrons with energy in the MeV range 
can produce an energy deposit in the TPC that 
mimics a WIMP signal. Nuclear recoils due to neutrons therefore must be efficiently identified and vetoed and this requires a detector to identify the neutron. The goal is to keep the neutron-induced background
 below about 0.1 events over a 200 t$\cdot$yr exposure, after all the software analysis cuts have been applied.\\
The Gd-PMMA both reduces and helps tag the neutron background. The high density of hydrogen in PMMA slows the neutrons to thermal velocities and the Gd captures the thermal neutrons resulting in a gamma cascade \cite{Yano:2017,gado:2019}, which is measured in a neutron veto detector surrounding the central TPC and/or in the TPC itself. Events from neutron capture have a different energy scale and hit pattern than single-scatter events from dark-matter interactions.
As Fig.~\ref{fig:DS_sketch} shows, eight 15 cm thick panels of \ce{Gd}-PMMA, each 3.5 m high and 1.6 m wide, form the lateral walls of the TPC volume.
Additional blocks (0.4 m by 0.4 m by 0.15 m each) of \ce{Gd}-PMMA  are mounted above the top and below the bottom of the TPC. None of these parts have a structural function: the walls must support their own weight, while the blocks at the top and bottom are supported and held in place by steel structures. For this reason there are no stringent requirements on the mechanical properties of the Gd-PMMA.
The thickness of the \ce{Gd}-PMMA parts, the gadolinium concentration and its uniformity are the three key requirements set
to achieve, given the design of DarkSide-20k, a neutron tagging inefficiency $\sim10^{-6}$. This number is obtained through a detailed Monte Carlo simulation based on the Geant4 package \cite{Geant4}, including  the full detector geometry and material composition, with energy thresholds of 200 keV and 50 keV in the Veto and TPC respectively.
The result is that the minimum necessary thickness is 15 cm, which comes from a trade-off between the need to moderate and capture the neutron and the need of detecting the capture $\gamma$-rays in the surrounding liquid argon. Given this requirement, the raw slabs before processing need to be about 17 cm, as reported in Tab.~\ref{tab:GdProp}. This is an important requirement for
developing the production procedure of \ce{Gd}-PMMA. Moreover, requiring that neutron capture on gadolinium dominates over neutron capture on hydrogen we found that the optimal concentration of gadolinium is between 0.5\% and 1\% in weight. Details on concentration and uniformity of gadolinium are given in Sec.~\ref{sec_conc}. As described in detail in Sec.~\ref{section:LabTests}, the final thickness of a sample is critical for our gadolinium mixing procedure and the subsequent polymerization steps, as the solidification time, and consequently, any
non-uniformity in the distribution of the gadolinium depends on the thickness of the final object.\\
Approximately 20 t of material are needed, before all the machining, to build the detector, thus the procedure
to produce the \ce{Gd}-PMMA must be scalable for industrial production.

\subsection{Gadolinium concentration and uniformity}
\label{sec_conc}
The guideline for establishing the concentration is to load the PMMA with an amount of gadolinium such that the
capture of neutrons by gadolinium dominates 
 over the same process on hydrogen, thus maximizing the detection probability of the neutrons. Considering the weighted average of all the seven \ce{Gd} isotopes,
the thermal neutron capture on gadolinium $\sigma_{Gd}$ is $\sim$4.9$\times$10$^4$ barns  \cite{Gdcapture}, while the thermal capture cross-section on hydrogen,
$\sigma_{H}$, is $\sim$3.3$\times$10$^{-1}$ barns \cite{Surto:2003}.
(The thermal neutron capture cross-sections on carbon and oxygen are orders of magnitude
lower than on hydrogen, thus their contribution to thermal neutron capture is negligible. ) \\
Given these cross-sections,
for the probability of a thermal neutron to capture on \ce{Gd} to be, as a reference,  $\sim$100 times greater than on \ce{H}, the \ce{Gd}  concentration
in mass needed is $\sim$1$\%$ with respect to the PMMA mass. A concentration between 0.5$\%$ and 1$\%$ still ensures that the capture on gadolinium is dominant, and this range of concentration allows a level of non-uniformity in the Gd distribution over a volume of 1 cm$^3$. Having a nominal concentration of 1$\%$ with a 50$\%$ non-uniformity means that in each 1 cm$^3$ there is at least 0.5$\%$ of Gd, so in every point of the material the capture on gadolinium is dominant. For this reason we set a uniformity requirement of 50$\%$. 
This is a critical aspect when considering the thickness of the samples, because the concentration in the vertical direction can be strongly affected by sedimentation during the solidification. We note the use of nanograins of 80-100 nm in diameter guarantees a substantially continuous distribution when compared with the thermalization length of the neutrons.\\
The above-mentioned Monte Carlo simulations confirm that a gadolinium concentration between 0.5$\%$ and 1$\%$  delivers comparable neutron tagging performance given the dominance of $\sigma_{Gd}$ over $\sigma_{H}$.

\subsection{Radiopurity requirements}\label{section:Radiopurity_req}
Requirements about radiopurity are particularly challenging and they are driven by the requirement that the concentration of \ce{U} and \ce{Th} in the \ce{Gd}-PMMA is such that the neutron background contribution of the Gd-PMMA material itself due to ($\alpha$, n) reactions is subdominant to the total background budget.
The $\gamma$ decays of radioactive contaminants may also generate signals in the same energy range as those produced by WIMPs. However, they are identified and rejected with high efficiency thanks to the different distributions in time and space of the argon scintillation light generated by an energy deposit due to a $\gamma$-ray or to a nuclear recoil (that is the hypothetical WIMP interaction) and to the corresponding features of the ratio between scintillation and ionization signals for the two categories of events \cite{bul:2006}.
The powerful pulse shape discrimination of argon-based detectors makes the $\gamma$ decays not a significant source of  background in the dark matter search.
 Still, the concentration of $\gamma$ contaminants must be kept under control to limit the rate of accidental coincidences in the detector that would spoil its performance by introducing a dead time for the WIMP search. An example is the accidental coincidence between a $\gamma$-ray induced signal in the neutron veto buffer due to the decay of a contaminant in the detector material (including the \ce{Gd}-PMMA) and a WIMP-like event in the TPC volume.\\
 The detailed quantitative requirements about the radiopurity depend on the specific design of DarkSide-20k and on a detailed background model, the description of which is beyond the scope of this paper. The output of the backgroud model however provides limits on the activities of the material components, which are reported in Tab.\ref{tab:GdProp}. The effort performed in screening all the components making the new hybrid material is of general interest in the search for rare events.

 \subsection{Summary of the requirements}
PMMA has been selected as the base material because it is rich in hydrogen and also because it can be produced with an excellent degree of radiopurity through a casting process, polymerizing its liquid monomer (methyl methacrylate, MMA).
Since there are no stable commercially available gadolinium compounds that are soluble in liquid MMA,
except for some particular very expensive complex compounds like Gadolinium acetylacetonate, we decided to make a dispersion of
gadolinium in the liquid monomer followed by the polymerization. We chose to use gadolinium oxide which is a stable and cheap compound, in the form of nanoparticles,
to maximize the uniformity of the \ce{Gd} distribution in the polymer.
The characterizations of the samples will mainly focus on verifying the fulfillment of all the mentioned requirements, that are summarized in Tab.~\ref{tab:GdProp}. From the point of view of the quality of the polymer and its mechanical properties, there are no stringent requirements, as the sheets of the neutron veto will not be subjected to particular mechanical stress, however, some key quantities were measured to verify that \ce{Gd}-PMMA has similar properties to pure PMMA, as will be shown in Sec.~\ref{sec:char}.

\begin{table*}[!h]
\begin{minipage}{\textwidth}
\begin{center}
\begin{tabular}{lc}
\hline
\hline
\textbf{Parameter} & \textbf{Value} \\
\hline
\hline
Gd concentration (weight)  & $0.5\%< \ce{Gd} < 1\% $\\
Gd homogeneity              & $\simeq\ 50 \%$ \\
Transparency of the hybrid Gd-PMMA material & not necessary\\
Machinable   & yes \\
Stable at 87 K &  yes\\
Thickness \footnote{\label{1sttablefoot} Before the machining of the final pieces.}         & $\sim$ 17 cm \\
Maximum size \footref{1sttablefoot}      & sheets of $\sim$ 4 m $\times$ 2 m \\
$^{238}$U, $^{235}$U, $^{232}$Th activity of Gd$_2$O$_3$
&  $<$ 20 mBq/kg \\
$\gamma$ contaminants activity of Gd$_2$O$_3$ 
& $<$ 2 mBq/kg  \\
Amount needed \footref{1sttablefoot}    & about 20 t \\
\hline
\end{tabular}
\end{center}
\caption{Target specifications driving the development of the \ce{Gd}-PMMA. 
The $^{238}$U, $^{235}$U,$^{232}$Th acceptable concentrations depend on multiple parameters since the background is a combination of values in different parts of the chains with their corresponding yields.} 
\label{tab:GdProp}
\end{minipage}
\end{table*}

%% file: Radiopurity_new.tex
\section{Radiopurity}
\label{sec:Radiopurity}
\subsection{Radiopurity of the ingredients}
\label{sec:Radio_ing}
The first step to produce the \ce{Gd}-loaded hybrid material with the necessary radiopurity is the selection of radiopure components. For this reason, the components of the hybrid material were subjected to radiopurity screening using the Inductively Coupled Plasma Mass Spectrometry (ICP-MS) technique or High-Purity Germanium detectors (HPGe).
The ICP-MS determines the concentration of an isotope in a sample by measuring mass/charge ratio~\cite{icpms}, while an HPGe detector consists of a semiconductor diode that detects traces of unstable isotopes thanks to 
$\gamma$-ray spectroscopy. Most of the assays done with HPGe detectors were carried out in a dedicated facility at the Canfranc Underground Laboratory (LSC), in a radon abated environment ~\cite{hpge}, while ICP-MS measurements were done at LNGS~\cite{lngs_ge}.\\
The two main ingredients of the \ce{Gd}-PMMA are MMA
(the monomer from which the PMMA is produced), and \ce{Gd2O3}. After a thorough search for suppliers, we have found that commercial gadolinium oxide \ce{Gd2O3},
delivered by the Shin-Etsu Chemical Co. Ltd. Company (Japan) has a suitable level of radioactive contaminants. We selected \ce{Gd2O3} in the form of nanoparticles with a diameter between 20 and 80 nm, with 3N purity, produced from lots GD-0BB-035 and GD-0BB-038. As reported
in Tab.~\ref{tab:gd2o3}, we screened with an HPGe detector three different samples
and they show very low contamination levels. In particular the third sample, produced from lot GD-0BB-038, has $^{238}$U, $^{235}$U $^{232}$Th and $^{40}$K contamination levels compatible with the background requirements reported in Tab.\ref{tab:GdProp}. 

\begin{table*}[h]
\begin{minipage}{\textwidth}
    \centering
    \begin{tabular}{lccc}
    \hline
    \hline
        \textbf{Isotope} & \textbf{\ce{Gd2O3} Shin-Etsu \#1} & \textbf{\ce{Gd2O3} Shin-Etsu \#2} & \textbf{\ce{Gd2O3} Shin-Etsu \#3} \\
                & [mBq/kg]          & [mBq/kg] & [mBq/kg] \\
        \hline \hline
         $^{238}$U-$^{234m}$Pa & <1.2$\cdot$10$^3$  & <637 & <99 \\
         $^{238}$U-$^{226}$Ra\footnote{\label{nota2} To be compared with the values in Tab.\ref{tab:GdProp}.}  & 13.6$\pm$3.0  & 6.6$\pm$1.8 & 2.68$\pm$0.47 \\
         $^{232}$Th-$^{228}$Ac & <30 & <24 & <6 \\
         $^{232}$Th-$^{228}$Th\footref{nota2} & <27 & <19 & 2.31$\pm$0.68  \\
         $^{235}$U-$^{235}$U & <51 & <25 & <1.5  \\
         $^{235}$U-$^{227}$Ac\footref{nota2} & <82 & <57 & <6.5  \\
         $^{40}$K & <37 & <24 & <13 \\
         $^{60}$Co & <2.5 & <1.3 & <0.62 \\
        $^{137}$Cs & <4.0 & <2.2 & <0.70\\
        $^{138}$La & <3.2 & <2.0 & <0.71\\
         $^{176}$Lu & 12.9$\pm$2.6 & 12.1$\pm$2.1 & 2.3$\pm$0.35\\
         \hline
    \end{tabular}
    \caption{Assay results of three \ce{Gd2O3} samples from the Shin-Etsu company, performed by the DarkSide Collaboration with a HPGe detector.}
    \label{tab:gd2o3}
    \end{minipage}
\end{table*}
Concerning MMA, data reported by other experiments show that MMA can both be delivered with excellent levels of radiopurity and contamination during the polymerization phase is avoidable
\cite{Deap:con,Juno:acr}, for these two reasons we did not screen the monomer used for the laboratory phase.\\
In the context of the industrial tests described in Sec.~\ref{sec:IndustrialTests}, an ICP-MS screening
was performed on the MMA provided by the selected industrial partner (Clax s.r.l., see Sec.~\ref{sec:IndustrialTests} for details). Measurement showed concentrations of thorium and uranium well below the mBq/kg,
as reported in Tab.~\ref{tab:ICPMS}. This is a preliminary result that indicates that MMA can generally be a sufficiently radiopure component, a detailed screening should then be carried out once the industrial partner for the final production has been identified.\\
\begin{table}[tb]
    \centering
    \begin{tabular}{ccc}
    \hline 
    \hline
         \textbf{Sample} &  \textbf{$^{232}$Th} [mBq/kg] & \textbf{$^{238}$U} [mBq/kg]\\
         \hline \hline
        Clax MMA & $<$0.041 & $<$0.12 \\
         \hline
    \end{tabular}
    \caption{ICP-MS measurement of the MMA from Clax s.r.l. company used for the industrial tests.}
    \label{tab:ICPMS}
\end{table}
In addition to MMA and gadolinium oxide, the hybrid material contains a non-ionic surfactant, Polyoxyethylene (5) nonylphenylether,
better known with the commercial name of Igepal CO-520\textsuperscript{\textregistered}. This is used to minimize the formation of nanoparticle aggregates,
increasing the uniformity of the oxide distribution, as detailed in Sec.~\ref{sec:surface_treat}.
Compounds such as Igepal CO-520\textsuperscript{\textregistered}
(nonylphenol ethoxylates) are often produced by the reaction of nonylphenol with ethylene oxide,
with the addition of potassium hydroxide as catalyst \cite{Report:potassio}. The screening carried out with HPGe detector shows, as reported in Tab.~\ref{tab:Igepal_tot}, that the amount of residual $^{40}$K present in the compound is indeed too high for the requirements of the experiment.
It was then necessary to develop a procedure, described in Sec.~\ref{sec:K_reduction}, to reduce it before its usage.

\begin{table}[tb]
    \centering
    \begin{tabular}{cccc}
    \hline 
    \hline
         \textbf{Sample} &  \textbf{$^{232}$Th} [mBq/kg] & \textbf{$^{238}$U} [mBq/kg] & \textbf{$^{40}$K} [mBq/kg]\\
          & (ICP-MS) & (ICP-MS) & (HPGe)\\
         \hline \hline
         Igepal CO-520 & $<$0.041 & $<$0.12 & (31.9$\pm3.2)\cdot$10$^{3}$ \\
         \hline
    \end{tabular}
    \caption{Surfactant screening results both from ICP-MS and HPGe. For the full results obtained with HPGe see Tab.~\ref{tab:HPGe_Ige} }
\label{tab:Igepal_tot}
\end{table}
Finally, during the production of Gd-PMMA, we used two additives necessary to initiate the polymerization, as we will further specify in Sec.~\ref{section:LabTests}, but since their concentration is of the order of ppm and a fraction of their mass is then released in the form of gas during the polymerization reaction, we considered acceptable not to do the screening.\\

\subsection{Potassium reduction in the surfactant}
\label{sec:K_reduction}
To reduce the $^{40}$K content, the strategy adopted was to perform a purification to remove potassium using an ion exchange resin, a well-known technique \cite{resin:book,resin:book2}.\\
In particular, we adopted the so-called "batch method", often used for non-ionic surfactants, especially with high viscosity,
as in the case of the Igepal CO-520\textsuperscript{\textregistered}. We used
AmberChrom\textsuperscript{\textregistered} 50WX4 from Merck Millipore, in the hydrogen form with a 100 mesh (CAS: 69011-20-7). It is a strong cationic resin that removes K$^+$ and substitutes it with H$^+$ \cite{resin:ginn}.\\ The purification strategy is conducted as follows:
\begin{enumerate}
    \item The resin is washed with ethanol three times until the solvent remains colorless after being in contact with the resin \cite{resin:ginn}.
    \item The resin is reactivated with 3M hydrochloric acid.
    \item The resin is washed with deionized water and partially dried under a fume hood.
    \item The resin, still wet, is immersed in the Igepal CO-520\textsuperscript{\textregistered}. The mixture is left under magnetic stirring.
    \item The grains of the resin are separated from the purified surfactant with a centrifugation process.
\end{enumerate}
Both the quantity of resin and the stirring time have been optimized during the tests, informed by several measurements of the potassium content. In particular, the purified Igepal CO-520\textsuperscript{\textregistered} was screened with Inductively Coupled Plasma Atomic Emission Spectrometry (ICP-AES). Some of the prepared solutions were also screened via ICP-MS, to crosscheck the measurements. The best purification procedure consists of using 10\%$_w$ of resin, with respect to surfactant mass, and keeping them under magnetic stirring for one week. 
One of the final samples, purified with this procedure, underwent HPGe screening, to evaluate the concentration of all the contaminant isotopes. Results are reported in Tab.~\ref{tab:HPGe_Ige} and it can be seen that the $^{40}$K activity was reduced by about a factor of 250. 
                     
\begin{table}[!h]
    \centering
    \begin{tabular}{lcc}
    \hline
    \hline
        \textbf{Isotope} & \textbf{Raw Igepal activity} & \textbf{Purified Igepal activity} \\
                & [mBq/kg]          & [mBq/kg] \\
        \hline \hline
        $^{235}$U & <51 & <9.4 \\
        $^{238}$U-$^{234m}$Pa & <4.8$\cdot$10$^3$  & <1.5$\cdot$10$^3$ \\
        $^{238}$U-$^{226}$Ra & <55 & <19 \\
        $^{232}$Th-$^{228}$Ac & <105 & <43 \\
        $^{232}$Th-$^{228}$Th & <43 & <12 \\
        $^{40}$K & (31.9$\pm$3.2)$\cdot$10$^3$ & 129$\pm$24 \\
        $^{137}$Cs & <28 & <4.4 \\
        $^{60}$Co & <23 & <3.3 \\ \hline
    \end{tabular}
    \caption{Results of the assay campaign conducted on both the raw and the purified surfactant. The screening was performed with a HPGe detector.}
    \label{tab:HPGe_Ige}
\end{table}

In parallel to the tests of the procedure with the ion exchange resin, we decided to reduce the amount of
surfactant used in the polymerization process. In the laboratory tests, we reduced up to a factor 100 the
Igepal CO-520\textsuperscript{\textregistered} amount, going from 1\%$_w$ with respect to the MMA mass, as used in our standard procedure, to 0.01\%$_w$. In the industrial tests, we opted for a surfactant concentration of 0.1\%$_w$, since the polymerization is more delicate.
This reduction is sufficient to achieve contamination values acceptable for the experiment.
Moreover, by combining the reduction in mass with the purification a considerable safety margin on the amount of potassium can be achieved.

%% file: Experimental_part_new.tex
\section{Laboratory production}
\label{section:LabTests}
As mentioned in Sec.~\ref{sec:Introduction}, \ce{Gd2O3} is not soluble in liquid MMA, so it is necessary to make a dispersion to mix them together.
The use of commercial \ce{Gd2O3} in the form of nanoparticles with a diameter between 20 and 80 nm,  maximizes the uniformity of the \ce{Gd} distribution in MMA.
As reported in Tab.~\ref {tab:GdProp}, a gadolinium concentration in mass between 0.5\%$_{w}$ and 1\%$_{w}$ is necessary for the final material.
The laboratory work was started aiming at obtaining a stable colloidal dispersion of gadolinium oxide in liquid MMA with a very small concentration (0.001\%$_{w}$ of \ce{Gd2O3}).
Then the concentration was progressively increased up to a maximum of 2\%$_{w}$ of gadolinium, corresponding to 2.3\%$_{w}$ of \ce{Gd2O3}, considering the molecular mass. The maximum concentration value tested was chosen to have a large safety margin, since it is twice the maximum required value.\\
To get a uniform distribution of \ce{Gd2O3} with a concentration up to 2.3\%$_{w}$, it is necessary to treat the nanoparticles in advance, to minimize the formation of aggregates and prevent their sedimentation.
It is indeed well known that nanoparticles tend to cluster and, consequently, deposit on the bottom of the mold during the polymerization process, spoiling the uniformity of the final material \cite{Aggregation}.
The basic idea is to treat the nanoparticle's surface to undergo a functionalization process.
Functionalization refers to the surface modification of nanoparticles, which includes the bonding
of chemicals or biomolecules onto the surface aimed at creating repulsive electrostatic forces and steric hindrance factors (that is the prevention or retardation of interaction as a result of a spatial structure of a molecule) \cite{Gdoxide_PMMA_2011,Colombo13}.
This is the first step of the developed procedure and it is performed by treating the nanoparticles with a commercial non-ionic surfactant in a non-aqueous solvent.
 This process introduces a repulsive force between the particles that stabilizes the dispersion.
 The second step consists in the polymerization phase, which has been optimized to produce thick (up to 22 cm) samples.
 In particular, the polymerization time depends on the sample thickness, since as the thickness increases, also the solidification time does,
 consequently a \ce{Gd2O3} deposit is more likely to form. To overcome this effect a procedure was required to fine-tune the polymerization temperature, the quantities of chemical initiators used, and the pre-polymerization phase.\\

\subsection{Surface treatment of the nanoparticles}
\label{sec:surface_treat}
As anticipated in Sec.~\ref{sec:Radio_ing}, to facilitate the dispersion of the Gd nanoparticles we chose to perform the functionalization with Polyoxyethylene (5) nonylphenylether,
branched (\ce{(C_2H_4O)_n \cdot C_{15}H_{24}O}), whose commercial name is Igepal CO-520\textsuperscript{\textregistered}, by Sigma Aldrich.
It is a non-ionic commercial surfactant that appears as a transparent and viscous liquid.
Gadolinium oxide and surfactant are added, in a 1:1 mass ratio (so both at 1\%$_w$ with respect to MMA), to a non-polar solvent, which is taken in a ratio of about 1:4 to the MMA, by volume. 2-butanone (Sigma Aldrich, $\ge99.0\%$) was chosen as the non-polar solvent due to its boiling temperature (79.6\celsius) which allows it to easily evaporate in the first minutes of the pre-polymerization phase.
Moreover, the boiling temperature is high enough to not create bubbles during the polymerization phase, if traces of 2-butanone remain in the sample after the pre-polymerization.
The mixture is then sonicated, in order to break up any agglomerate of nanoparticles and to favor their surface covering with the surfactant.
This procedure is carried out for 15 minutes in continuous mode, using a Sonic Ruptor 400 Ultrasonic Homogenizer, by Omni. \\
As anticipated in Sec.~\ref{sec:K_reduction}, several tests have been performed to reduce the amount of surfactant, in order to minimize the radioactive contamination.
The described steps have been performed both with 0.1\%$_{w}$ and 0.01\%$_{w}$ of Igepal CO-520\textsuperscript{\textregistered} with respect to the monomer.
These samples were characterized to evaluate the uniformity of the Gd distribution even in the presence of a lower quantity of surfactant
(see Tab.~\ref{tab:calcinations} in appendix \ref{Appendix}). Based on the results of the characterization, we can state that the procedure is successful also
using 0.01\%$_{w}$ of surfactant.

\subsection{Polymerization}
Since methylmethacrylate is a non-viscous liquid, during the polymerization phase clusters of nanoparticles form, and consequently an oxide deposit is created at the bottom of the mold. This happens even with treated nanoparticles because of the long polymerization time: the dispersion of treated gadolinium oxide remains stable for roughly one hour (see Sec.~\ref{sec:char}), while the full polymerization takes up to 24 hours.\\
Therefore, to minimize the formation of a \ce{Gd2O3} deposit, we perform the polymerization in two steps \cite{Colombo13}:
first, we start a pre-polymerization phase, during which the MMA viscosity increases due to the formation of a pre-polymer (or oligomer), and then the actual polymerization occurs, leading to the final solid polymer.\\
As a preliminary operation, the MMA is passed through a separation column containing aluminum oxide (Al$_2$O$_3$), to remove hydroquinone monomethyl ether (MeHQ), which is used as the stabilizer and polymerization inhibitor. Once the MMA is filtered, the sonicated dispersion of gadolinium oxide (prepared as described in Sec.~\ref{sec:surface_treat}) is added. The mixture is then heated on a plate and frequently
stirred manually. During this phase, the 2-butanone solvent starts to evaporate.
When the temperature reaches 80\celsius, 100 ppm of primary polymerization initiator, with respect to the MMA mass, is added. AIBN (2,2-Azobis (2-methylpropionitrile \ce{(CH_3)_2C(CN)N=NC(CH_3)_2CN})) was used as initiator. When the temperature is between  94\celsius\ and 100\celsius, a very vigorous boiling begins due both to the reaching of the boiling point of the monomer, and to the splitting of the AIBN, which produces gaseous nitrogen. At this point the viscosity of the mixture rapidly increases, because the initiator causes
the start of the radical polymerization reaction. The heating is stopped after 9 minutes
of boiling when a homogeneous whitish viscous mixture is obtained. At this point, the 2-butanone solvent is fully evaporated. \\
The second phase of the process starts with removing the beaker from the heating plate. Then, 600 ppm (with respect to the MMA mass) of secondary initiator are added, namely Lauroyl peroxide 98\% (\ce{[CH3(CH2)10CO]2O2}),
whose commercial name from Sigma Aldrich is Luperox. The whole compound is mixed manually, to dissolve the Luperox
and finally, it is transferred in a glass mold and placed in the oven at 55\celsius, for 24 hours. Consequently, the sample is completely solid.\\

\subsection{Samples produced at the laboratory scale}

\begin{figure}[]
	\centering
	\includegraphics[width=.45\textwidth]{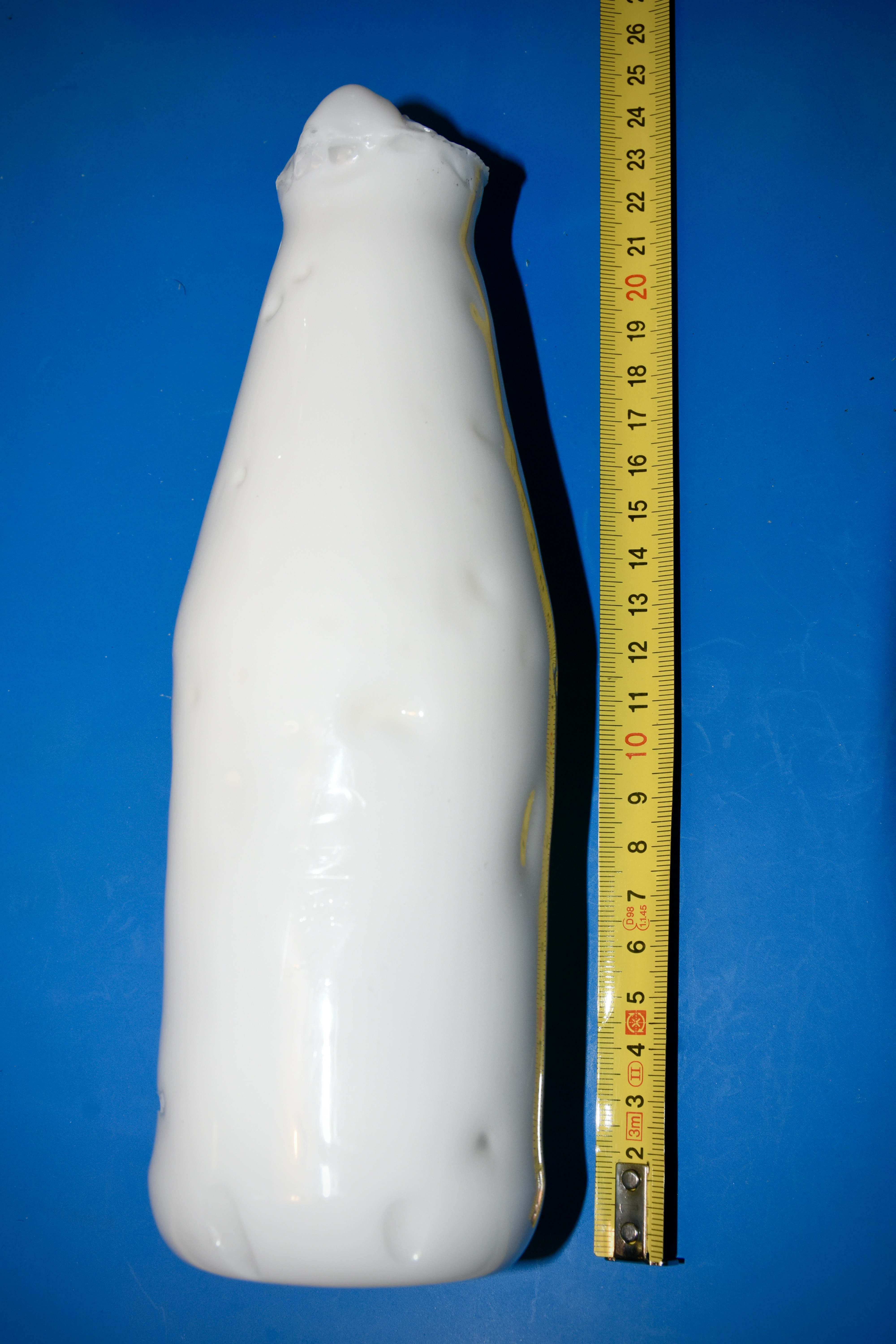}
	\caption{Sample 22 cm high, obtained starting from 600 g of MMA, loaded with 2\%$_w$ of \ce{Gd} (equal to 2.3\%$_w$ \ce{Gd2O3}).}
	\label{fig:sughino}
\end{figure}

Numerous samples were made, varying some of the parameters described above.
The gadolinium oxide and the surfactant concentrations were progressively modified throughout the optimization of the procedure, reaching 2.3$\%_w$ for the \ce{Gd2O3} and 0.01$\%_w$ for the Igepal CO-520\textsuperscript{\textregistered}. \\
Another key aspect was the optimization of the procedure to maximize the thickness of the samples, to reach the 17 cm required.
It was possible to produce samples with a thickness greater than 20 cm, as the one shown in Fig.~\ref{fig:sughino} that, in particular,
was produced with a 2.3\%$_w$ concentration of \ce{Gd2O3} and 2.3\%$_w$ of Igepal CO- 520\textsuperscript{\textregistered}.
To obtain a sample with these characteristics, the mass of MMA was progressively increased in the various productions,
also varying the other parameters of the procedure.


%% file: Characterization_new.tex
\section{Characterization}
\label{sec:char}
During the laboratory test phase, different aspects of the produced samples have been characterized. The procedures and the techniques described in this section were then applied to verify the quality of the industrial-scale samples (Sec.~\ref{sec:IndustrialTests}). \\
In particular, we report on the following aspects:
\begin{enumerate}
    \item Characterization of the surface treatment (Sec.~ \ref{sec:superf});
      \item State of the polymeric matrix (Sec.~ \ref{sec:Tg});
       \item Homogeneity of the \ce{Gd2O3} distribution (Sec.~ \ref{sec:Calcinations});
       \item Mechanical tests (Sec.~ \ref{sec:Mech_prop}).
   \end{enumerate}

\subsection{Characterization of the surface treatment}
\label{sec:superf}
\begin{figure}[!h]
	\centering
	\includegraphics[width=.8\textwidth]{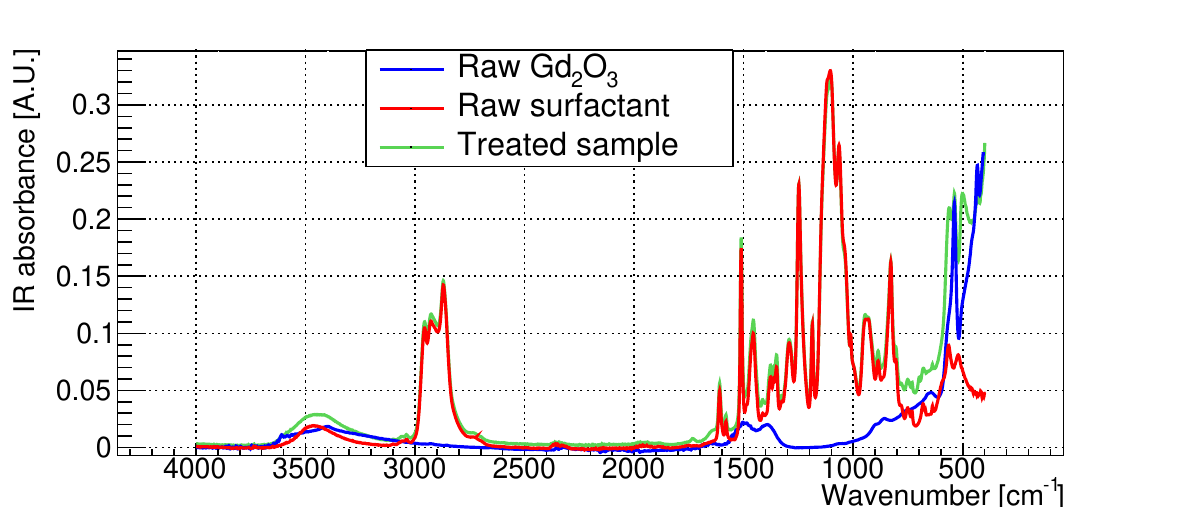} 
	\caption{IR spectra of the pure gadolinium oxide (blue), the pure surfactant (Igepal CO-520, orange), and the treated sample (green).}
	\label{fig:IR}
\end{figure}

The efficiency of the nanoparticles surface treatment procedure has been evaluated through two different types of measurements: 
the Fourier-Transform Infrared Spectroscopy (FTIR) \cite{FTIR} and Dynamic Light Scattering (DLS) \cite{DLS}.
The FTIR technique allows the identification of chemical substances or
functional groups present in the sample. Our measurements have been performed using the Alpha~II compact spectrometer by Bruker, in a wavenumber range from 4000~cm$^{-1}$ to 500~cm$^{-1}$.
The functionalization is singled out by comparing the IR absorption spectra of
the treated nanoparticles - after the functionalization procedure - with that of the 
raw \ce{Gd2O3} and raw surfactant. The treated samples were always in the form of dried powder, while the raw surfactant was in liquid form.
Fig.~\ref{fig:IR} shows the infrared spectra of the two raw ingredients superimposed on the spectrum of a functionalized sample. 
The nanoparticles covering is evidenced by the presence, in the spectrum of the treated sample, of the characteristic peaks
of the surfactant (the region between 700 and 1500 cm$^{- 1}$
and between 2700 and 3000 cm$^{- 1}$) in addition to those of the gadolinium oxide
(as can be seen in the region around 500 cm$^{- 1}$). This result can be considered as a first qualitative indication of the presence
of the surfactant on the nanoparticles.\\
Additional information has been obtained with the DLS technique, used to analyze the colloidal stability of the treated Gd nanoparticles in liquid.
The basic idea is to perform various DLS measurements as a function of time, to investigate the hydrodynamic size of particles (or clusters of nanoparticles) that have remained in suspension after a certain time. The variation of the particle hydrodynamic size distribution over time can give an indication of their sedimentation.
The DLS technique is based on the study of the Brownian motion of dispersed particles, from which it is possible to determine
their hydrodynamic diameter by measuring their speed thanks to the Stokes-Einstein equation.
The instrument used is the Zetasizer nano ZS90 by Malvern Panalytical. The light source used is a 10~mW 632.8 nm He-Ne laser, with an optical detector at 90\degree.\\
After the sample was prepared as described in Sec.~\ref{sec:surface_treat}, the suspension - just at the end of sonication - is diluted with pure 2-butanone, reaching a concentration of about 0.4 mg/ml.
However, in 2-butanone, and in general in low-viscosity solvents, it was not possible to obtain a fully stable dispersion on which to carry out a quantitative analysis. 
Although the dispersion is not entirely stable, the role of the surfactant is still crucial and to highlight the effect caused by its presence, it was decided to perform a DLS measurement by comparing a treated sample with an untreated one, i.e. free of surfactant. Moreover, in order not to alter the dispersion, the sample was left inside the instrument for the entire duration of the measurement. The two analyzed dispersions were obtained applying the exact same procedure (see Sec. \ref{sec:surface_treat}), but avoiding the addition of the Igepal CO-520\textsuperscript{\textregistered} in one of the two. \\
\begin{figure}[t]
    \centering
    \includegraphics[width=.8\textwidth]{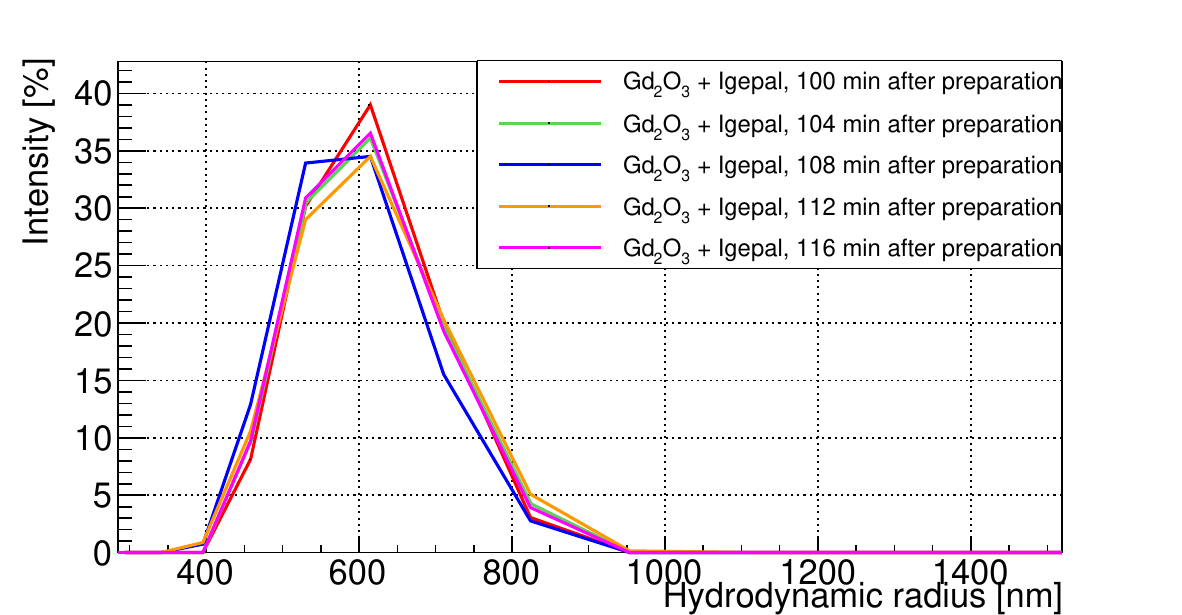}
    \caption{DLS measurements performed on a sample carrying an equal concentration of surfactant and \ce{Gd2O3} nanoparticles, diluted to a concentration of 0.4 mg/ml. The different curves were acquired starting 100 minutes after the preparation of the sample, at 4-minute intervals. The plot shows that in the 20-minute duration of the measurement, the population of nanoparticles with a hydrodynamic radius peaked at 600 nm is well reproducible, and therefore the suspension is stable.}
    \label{fig:DLS_curve}
 \end{figure}
The results obtained with the bare \ce{Gd2O3} nanoparticles were unsatisfactory, since the data did not even meet the basic quality requirements of the software of the instrument. This means that the suspension is not stable even on time scales of 15-20 minutes, which is the time required to complete the measurement, and very fast sedimentation of the nanoparticles occurs. 
On the other hand, the functionalized sample showed results that satisfied the data-quality controls, i.e. the stability of the suspension is compatible with the characteristic time scales of the measurement. So, it was possible to perform several measurements at different times after the end of the sonication. Analyzing the particle hydrodynamic size distribution over time, there is evidence of sedimentation, so the suspension with the surfactant is not completely stable, as mentioned previously. However, 100 minutes after the end of the sonication, there is still a clear population of nanoparticles dispersed, with a hydrodynamic radius of around 600 nm, as reported in Fig.~\ref{fig:DLS_curve}.  
In conclusion, the surface treatment does not allow the production of a fully stable suspension but reduces the clusterization and, therefore, the sedimentation. The results obtained with the functionalized nanoparticles are greatly improved with respect to the ones of the unfunctionalized \ce{Gd2O3}. \\
The functionalization procedure, combined with the developed polymerization procedure, allows to obtain \ce{Gd}-PMMA samples with a good nanoparticles distribution (as it is evidenced from the characterization in Sec.~\ref{sec:Calcinations}).

\subsection{State of the polymeric matrix}
\label{sec:Tg}
The effect of the presence of the nanoparticles in the polymeric matrix has been evaluated by looking for possible differences in the glass transition temperature (T$_g$), which is one of the most important parameters to evaluate the thermomechanical properties of a polymer \cite{Tg_generic}. 
The T$_g$ value is obtained through  Differential Scanning Calorimetry (DSC) using a DSC1 Star$^{e}$ System by Mettler Toledo. At the glass transition temperature, there is a variation of the specific heat, detectable from the DSC curve. In particular, 
we have adopted the most widely used definition, according to which the midpoint T$_g$ is defined as the point at which the first derivative of the DSC curve is maximum.\\
Three thermal ramps were carried out: the first, in heating, from room temperature to 250\celsius, the second, for cooling,
from 250\celsius\ down to -50\celsius\ and the last, a second heating phase, from -50\celsius\ to 250\celsius.
The glass transition temperature found for all samples, prepared with an identical procedure, is between 116\celsius\ and 119\celsius. Considering the non-negligible uncertainty associated with the DSC technique, the values are considered fully compatible \cite{DSC_rep}.
This means that the developed procedure leads to a hybrid material with reproducible properties from the point of view of the polymer matrix.
Moreover, the presence of \ce{Gd2O3} does not cause significant variations of the T$_g$ with respect to the pure PMMA reference values \cite{PMMA_Tg, Tg_poli}.

\subsection{Homogeneity of the \ce{Gd2O3} distribution}
\label{sec:Calcinations}
\begin{figure}[!h]
     \centering
     \includegraphics[width=0.6\textwidth]{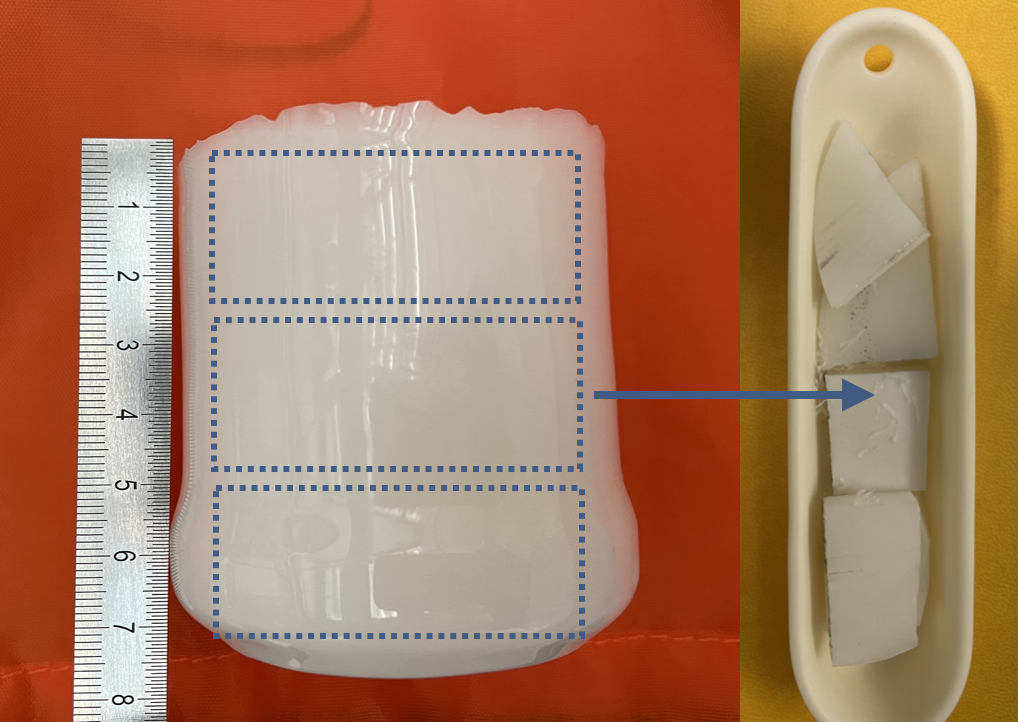}
     \caption{Scheme of samples sectioning for the calcination tests.}
     \label{fig:Sec_calc}
 \end{figure}

\begin{figure}[!h]
    \centering
    \includegraphics[width=\textwidth]{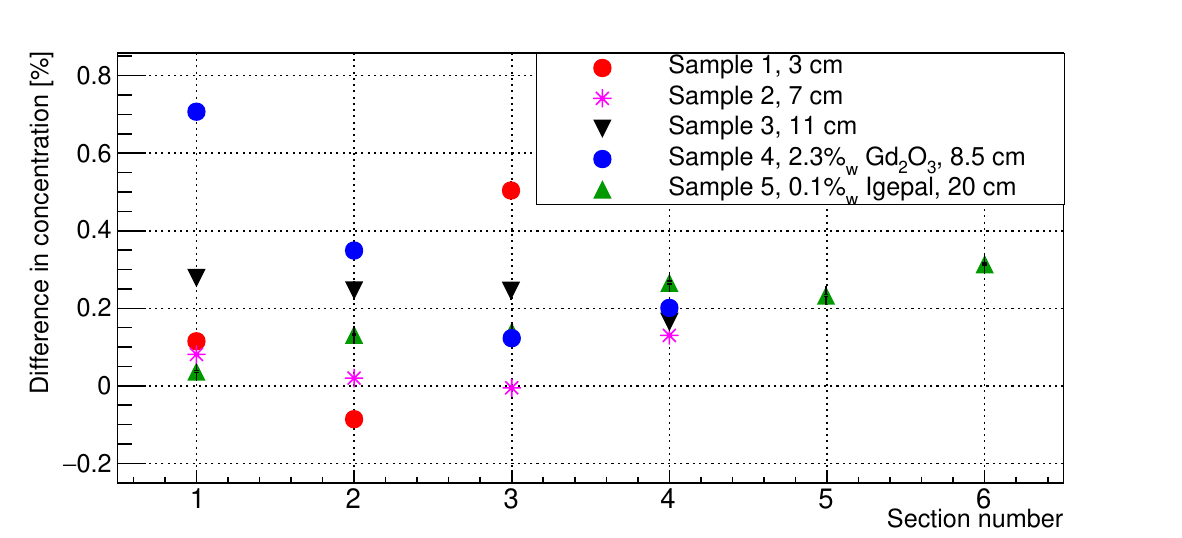}
    \caption{Results of the most representative calcination tests. On the y-axis, $\Delta C$ is the difference between the measured and the nominal mass concentration (expressed in [\%$_{w}$]) values. Referring to Tab.\ref{tab:calcinations}, the plot shows the difference between the fourth and the third columns for samples: 19-05-2, 26-05-2, 29-09-1, 05-10-1 and 09-12-1, here called 1,2,3,4 and 5 respectively. In all displayed plots, the y errors are covered by the markers.}
    \label{fig:Calc_lab}
\end{figure}

To characterize the homogeneity of the \ce{Gd2O3} distribution along the whole height we have analyzed portions of the same sample, taken at different heights starting from the bottom of the mold, as illustrated in Fig.~\ref{fig:Sec_calc}.
The measured samples have been divided into several sections, depending on the height of the sample itself (usually the highest part, the central part, and the lowest part). Subsequently, fragments of each section have been weighed with a precision balance, and placed in an alumina crucible, which in turn was placed in a tubular oven open at the ends, to facilitate the escape of gases. The sample was calcinated, i.e. heated up to 600\celsius, to remove all the organic substances, through thermal decomposition \cite{Calcination}. In particular, the standard thermal cycle consisted of initial heating from room temperature to 300\celsius, followed by a second heating ramp up to 430\celsius, and, finally, the sample was brought to 600\celsius. Each of these temperatures was kept constant for 20 minutes. Subsequently, the sample was slowly cooled to 250\celsius\ and, finally, to room temperature.
At the end of cooling in the crucible only inorganic residues are present. In our case the residue consists only of the gadolinium oxide initially present in the portion of the sample analyzed. By weighing the residue and comparing the result with the initial mass, a measurement of the \ce{Gd2O3} concentration ($C_{\textnormal{Gd}_2\textnormal{O}_3}$) by weight is obtained, following:
\begin{equation*}
    C_{\textnormal{Gd}_2\textnormal{O}_3} = \frac{100\cdot m_{\textnormal{Gd}_2\textnormal{O}_3}}{m_{\textnormal{Gd-PMMA}}}\,\, ,
\end{equation*} 
where $m_{\textnormal{Gd}_2\textnormal{O}_3}$ is the mass of the $\textnormal{Gd}_2\textnormal{O}_3$ residue after the thermal cycle, and $m_{\textnormal{Gd-PMMA}}$ is the mass of the starting Gd-PMMA sample portion subjected to this analysis.
By analyzing the different sections, the uniformity of the distribution of \ce{Gd2O3} along the vertical direction of the sample is obtained. The most significant results are reported in Fig.~\ref{fig:Calc_lab}. In this figure, the y-axis reports a concentration difference $\Delta C$, which expresses the difference between the measured Gd$_2$O$_3$ concentration and the nominal one. \\
Samples 1, 2, and 3 carry a concentration equal to 1\%$_w$ of \ce{Gd2O3} and an equal concentration of surfactant, they are respectively  3 cm, 7 cm, and 11 cm thick. Sample 4 is 8.5 cm thick, and has a \ce{Gd2O3} and Igepal concentration of 2.3\%$_w$ and 2\%$_w$, respectively. Finally, Sample 5 is 20 cm high and carries a concentration of 1\%$_w$ \ce{Gd2O3} and 0.1\%$_w$ Igepal, it was obtained following the final procedure.
Thanks to the procedure used and the thermal cycle inside the muffle furnace, we do not expect any external contamination that could have increased the inorganic residue. However to study any systematic effects and verify the presence of inorganic residues other than gadolinium oxide, we also performed the same thermal treatment on a pure PMMA sample. We did not find any residual component, the difference between the initial and final masses is zero, within the error margin of the balance. \\
We can therefore conclude that the procedure described in the previous sections leads to homogeneous samples. With the exception of the lower section of sample 4, where there is a greater deposit, all the samples shown in Fig.~\ref{fig:Calc_lab} have an uniformity within 50\%, as required. It is useful to observe that the concentration variations are always greater than zero, this is due to the evaporation of a few grams of MMA during the polymerization phases.\\ 
In Tab.~\ref{tab:calcinations} in appendix~\ref{Appendix}, calcination results obtained on a wider range of laboratory samples are reported. Those tests have been done in order to investigate the reproducibility of the homogeneity results,  both increasing the height of the samples and the \ce{Gd2O3} concentration.

\subsection{Mechanical tests}
\label{sec:Mech_prop}
\begin{figure}[!h]
    \centering
    \includegraphics[width=.8\textwidth]{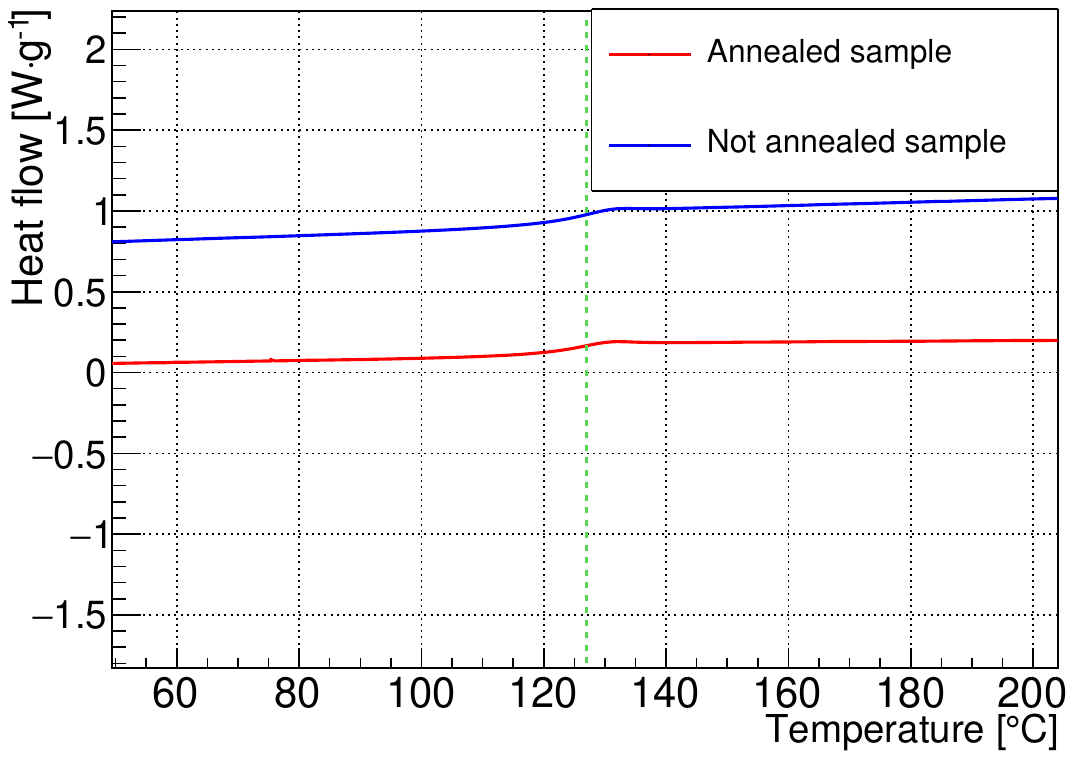}
    \caption{DSC curves of two specimens derived from the same sample, one annealed (blue) and the other not (red). The green dotted line is found in correspondence with the T$_g$ value for both specimens (around 127\celsius), which has not changed as a result of the thermal cycles.}
    \label{fig:DSC_annealing}
\end{figure}

\begin{figure}[!h]
    \centering
    \includegraphics[width=.8\textwidth]{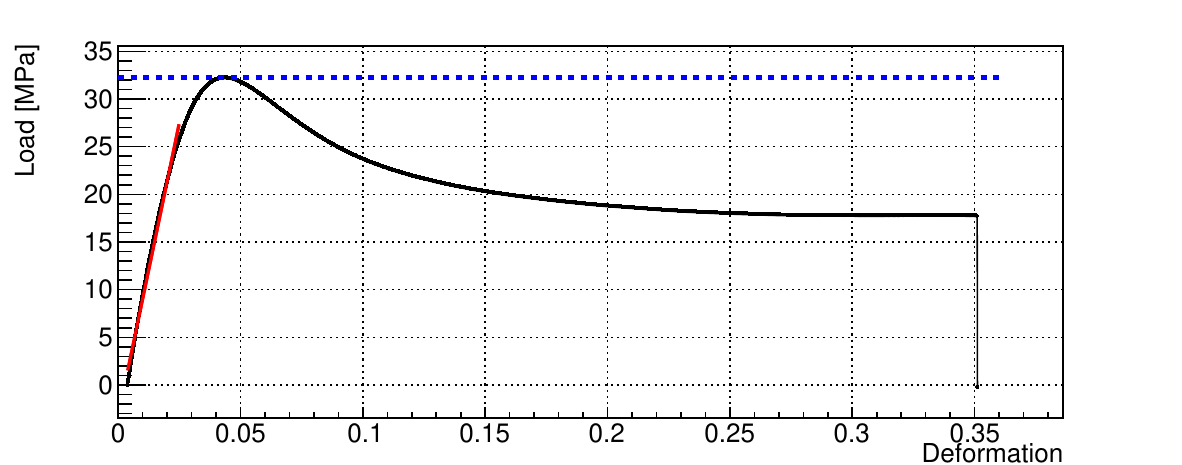}
    \caption{Stress-strain curve of a sample subjected to a cooling cycle in LN$_2$. The deformation is adimensional, since it is defined as $\Delta $L/L. The red line was used to fit the stress-strain curve in its elastic region for the calculation of the Young modulus, while the blue dotted line underlines to the yield stress value for this sample.}
    \label{fig:YM_ours}
\end{figure}

As anticipated in Sec.~\ref{sec:requirements} we used the characteristic values of pure PMMA as a benchmark to which we compared some mechanical properties of Gd-PMMA.
It was decided to measure, among all the mechanical properties, Young's modulus and the tensile strength because they are often used as key basic properties and also because they can be measured through simple and quick tensile tests.\\
Since for the DarkSide-20k experiment, operation at temperatures around 87 K is required, it was also necessary to cool the samples to cryogenic temperature and check their integrity. 
In addition, after a visual inspection, it was important to repeat the measurement of the mechanical properties to estimate any possible effects due to the cooling cycles.\\
Before performing any cool down or mechanical test, we annealed the samples to reduce all the residual stresses possibly accumulated during the polymerization and the mechanical machining. The thermal cycle was carried out in an oven, which can be either static or ventilated, and consisted of three phases: the heating ramp, the isothermal phase, and the cooling ramp. During the first phase, the temperature is increased, starting from room temperature, by 20\celsius\ every hour, until reaching 85\celsius. This temperature - close to T$_{g}$, but still more than 20\celsius\ below - is then maintained for 20 hours. Finally, a very slow cooling phase starts: temperature is decreased by 5\celsius\ every hour, down to room temperature. This procedure does not alter the properties of the polymer: this has been proven by comparing the glass transition temperature of annealed and not annealed portions of the same samples (see the DSC curves reported in Fig.~\ref{fig:DSC_annealing}). Note that the T$_{g}$ of this sample is different from the one reported in Sec.~\ref{sec:Tg}, but still in the range of pure PMMA. 
The samples, after the annealing process, were cooled down to 77 K by immersing them in liquid nitrogen. Since the Gd-PMMA will be used in liquid argon, i.e. at 87 K, the test in liquid nitrogen is conservative. The cooling from room temperature was performed using cold gas nitrogen during the first phase and then gradually submerging the sample with liquid nitrogen. 
Throughout the process, temperatures in different points of the sample were monitored - using Pt-100 sensors - and cooling was adjusted so that the maximum temperature gradient did not exceed 50\celsius. Samples of various thicknesses (from 1 mm to 20 cm) have been cooled and have been maintained in LN$_2$ from a few hours up to 2 weeks. 
None of the samples showed any sign of damage or degradation when visually inspected after the thermal cycle. In some cases, after the annealing step and a complete thermal cycle, the samples were subjected to a tensile test. One of the results is reported in Fig.~\ref{fig:YM_ours}.
The Young's modulus of this sample is 1.3 $\pm$ 0.1 GPa, obtained from the fit of the curve in the elastic region, while the ultimate tensile strength is 32.28 $\pm$ 0.04 MPa.
Young's modulus values of samples subjected to annealing and a full thermal cycle do not show significant variations compared to the values of pure PMMA usually found in the literature, which are between 0.6 GPa and 3.3 GPa \cite{Young modulus, CTE_PMMA}. Therefore we can conclude that neither the presence of nanoparticles in the polymeric matrix, nor the treatments undergone significantly modify the mechanical properties compared to pure PMMA.\\

%% file: Industrial_tests_new.tex
\section{Industrial scale tests}
\label{sec:IndustrialTests}
\begin{figure}
    \centering
    \includegraphics[width=.8\textwidth]{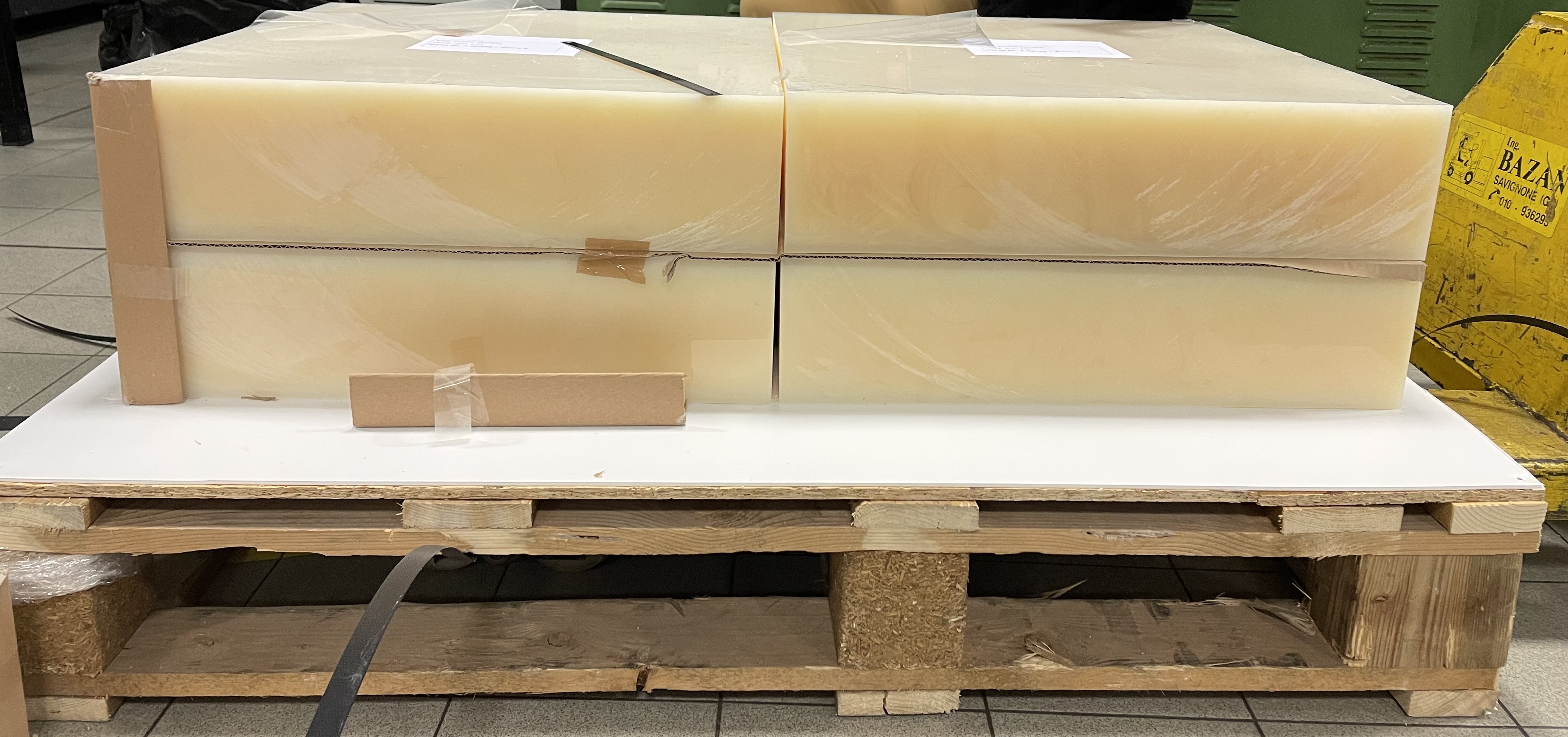}
    \caption{ Picture of four 12 cm thick sheets produced at Clax Italia s.r.l. }
    \label{fig:lastre}
\end{figure}

Since the amount of material required for the DarkSide-20k experiment is about 20 t, an important part of the activity consisted in transferring the procedure - validated at the laboratory scale - to an industrial production line.
We chose as partner an Italian company, Clax Italia s.r.l. \cite{Clax}. The transfer of the technology required an optimization of the mixing and polymerization procedures to adapt them to the production of large masses (in the order of tens of kilograms) and to the infrastructures available at the company. The key aspect on which the technology transfer work was concentrated was the viscosity setting to maximize the homogeneity of the gadolinium distribution. \\
The final industrial procedure involves a first mixing phase when the \ce{Gd2O3} nanoparticles and the Igepal CO-520\textregistered\ are directly incorporated in a small
quantity of MMA and subjected to vigorous mechanical stirring. Then this mixture is added to the final amount of MMA and the pre-polymerization stage is performed in a dedicated reactor, which consists of a heated container with a mechanical stirrer. During this phase the viscosity increases: as said, various tests were carried out, varying the duration and heating during this phase, reaching different viscosity values, as shown in Tab.~\ref{tab:clax_samp}.
Finally, the high-viscosity syrup is injected into a glass mold which is kept under high pressure during the polymerization in an autoclave. The polymerization time depends on the thickness of the sample: for the 12 cm thick sheets the full thermal cycle in autoclave has lasted 12 days.
Several large-scale samples were produced, varying not only the viscosity but also the amount and type of additives. Some of the 12 cm thick sheets produced are shown in Fig.~\ref{fig:lastre} Entering into the details of each procedure is beyond the scope of this work, but a summary of the main parameters is presented in Tab.~\ref{tab:clax_samp}.
\begin{table}[!h]
\centering
\begin{tabular}{ccccccccl}
\hline
\hline
\begin{tabular}[c]{@{}c@{}}\textbf{Sample}\\\textbf{name}\end{tabular} & \begin{tabular}[c]{@{}c@{}}\textbf{Surface}\\{[}cm$^2$]\end{tabular} & \begin{tabular}[c]{@{}c@{}}\textbf{Thick.}\\{[}cm]\end{tabular} & \begin{tabular}[c]{@{}c@{}}\textbf{Gd}$_2$\textbf{O}$_3$\\{[}\%$_w$]\end{tabular} & \begin{tabular}[c]{@{}c@{}}\textbf{Igepal}\\{[}\%$_w$]\end{tabular} & \begin{tabular}[c]{@{}c@{}}\textbf{T}\\{[}\celsius]\end{tabular} & \begin{tabular}[c]{@{}c@{}}\textbf{P}\\{[}bar]\end{tabular} & \begin{tabular}[c]{@{}c@{}}\textbf{Solv.}\\{[}y/n]\end{tabular} & \begin{tabular}[c]{@{}l@{}}\textbf{Visc.}\\{[}cP]\end{tabular} \\
\hline \hline
Clax 1-1 & 40$\times$40 & 4 & 1 & 1 & Std & 1 & yes & 400 \\
Clax 1-2 & 40$\times$40 & 4 & 1 & 1 & Std & 1 & yes & 400 \\
Clax 1-3 & 40$\times$40 & 4 & 1 & 1 & Std & 1 & no & 400 \\
Clax 2-1 & 50$\times$50 & 12 & 1 & 0.1 & $\approx$50 & 10 & no & 1000 \\
Clax 2-2 & 50$\times$50 & 12 & 1 & 0.1 & $\approx$50 & 10 & no & 1000 \\
Clax 2-3 & 50$\times$50 & 12 & 1 & 0.1 & $\approx$50 & 10 & no & 1000 \\
Clax 2-4 & 50$\times$50 & 12 & 0.80 & 0.1 & $\approx$50 & 10 & no & 400 \\
Clax 2-5 & 50$\times$50 & 12 & 0.85 & 0.1 & $\approx$50 & 10 & no & 400 \\ \hline
\end{tabular}
  \caption{An overview of the samples produced at Clax Italia s.r.l.. In addition to the variations of the parameters shown in the table, the procedures involved other small modifications from one sample to the other (addition of cross-linking agents, chain transfer agents, etc.). The column "Solv." indicates if the solvent was used or not, the column "Visc." indicates the value of the viscosity in the syrup. "Std" means that the company followed its standard thermal cycle used for pure PMMA, the details were not shared with us. }
\label{tab:clax_samp}
\end{table} 
Note that the use of 2-butanone as a solvent for the preparation of the suspension, which was adopted in the first three industrial samples, was then removed, further simplifying the procedure.
Thanks to the use of a dedicated reactor, both the mixture was kept continuously under mechanical stirring, and the viscosity was increased during the pre-polymerization phase, up to 1000 cP, thus making it possible not to use a liquid solvent to ensure the uniformity of the nanoparticles distribution.

\subsection{Industrial samples characterizations}
\begin{figure}[!h]
    \centering
    \includegraphics[width=.8\textwidth]{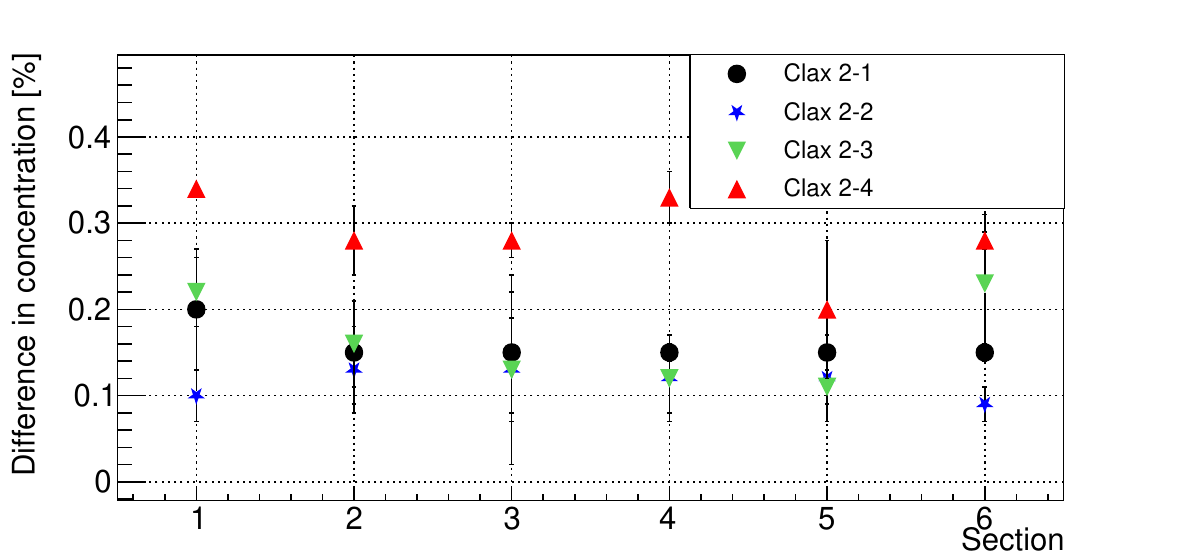}
    \caption{Deviation from the nominal value of the \ce{Gd2O3} concentration in six sections, taken at different heights. The results for four different sheets are reported.}
    \label{fig:Calc_Clax}
\end{figure}

The characterization of industrial samples was initially focused on the uniformity of the nanoparticles distribution. Following the procedures developed during the laboratory test phase, the homogeneity was measured by exploiting the calcination technique. The sheets have been cut in six sections at different heights with respect to the bottom of the mold. Small fragments from each section were subjected to the thermal cycle described in Sec.~\ref{sec:Calcinations}, then the inorganic residue, corresponding to the \ce{Gd2O3} was precisely weighed. We report in Fig.~\ref{fig:Calc_Clax} the results obtained with the 12 cm thick sheets produced. They show that the industrial procedure was successful in producing a sheet with a thickness that is 70\% of the one needed for the DarkSide-20k experiment. The results of the uniformity of the gadolinium oxide distribution, give an indication that the same procedure can lead to satisfactory results even with thicker plates. Furthermore, since there is still room to increase the polymerization speed and tune the viscosity of the pre-polymer, we are confident to obtain sheets of 17 cm thickness. 

Based on the results of the measurement of the uniformity of \ce{Gd2O3} and of the visual inspection aimed at
identifying any macroscopic defects in the samples, the sheet named "Clax 2-3" was selected as the best candidate. On the latter, further characterizations were then carried out, following what was done for the laboratory scale samples.
\begin{figure}[!h]
    \centering
    \includegraphics[width=.8\textwidth]{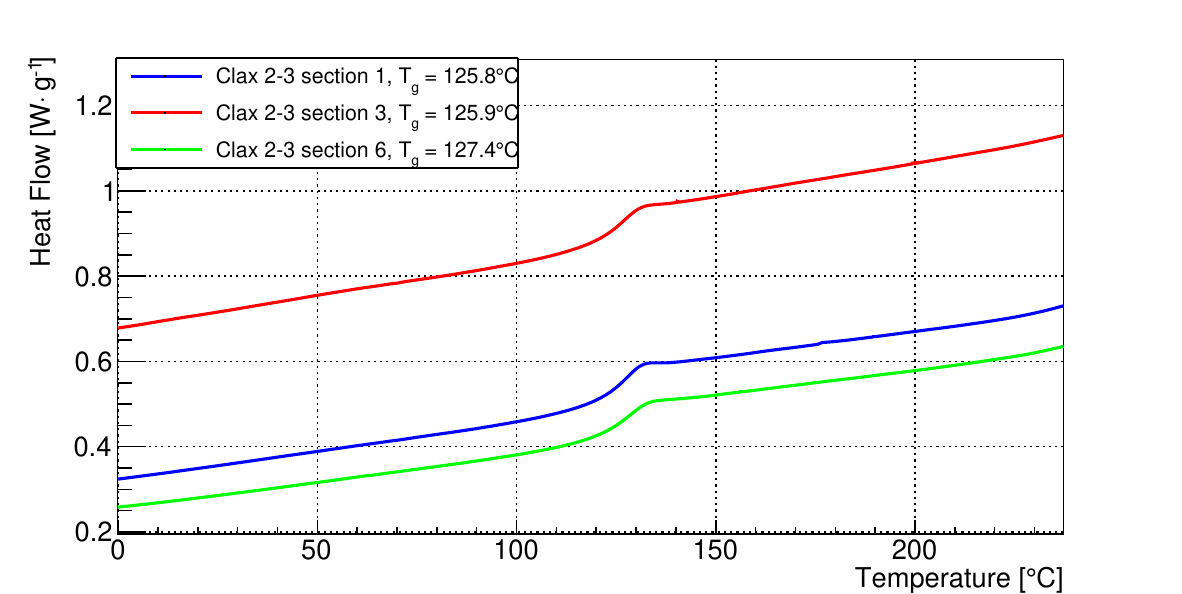}
    \caption{DSC curves different sections of the sample Clax 2-3.
    The reported T$_g$s fall within the systematic uncertainty equal to $\pm$2.5\celsius\ obtained from reproducibility measurements carried out on a single sample of Gd-PMMA.
    The T$_g$ is thus compatible along the sample’s thickness.}
    \label{fig:Clax_DSC}
\end{figure}
For what concerns the glass transition temperature, results are reported in Fig.~\ref{fig:Clax_DSC} and it can be seen that the value is uniform across the sheet and compatible with the one of pure PMMA.
The mechanical properties were investigated with tensile tests and thermomechanical tests, in order to derive the coefficient of thermal contraction. As for the tensile tests, the values of Young's modulus and ultimate tensile strength are reported in Tab.~\ref{tab:Clax YM}.
\begin{table}[!h]
    \centering
    \begin{tabular}{ccc}
    \hline
    \hline
        \textbf{Specimen} & \textbf{Young Modulus} & \textbf{Ultimate tensile strength} \\
         & [GPa] & [MPa] \\
         \hline \hline
        Clax 2-3 a & 1.4	$\pm$ 0.2 & 44.05 $\pm$ 0.05 \\
        Clax 2-3 b & 1.2	$\pm$ 0.2 & 39.29 $\pm$ 0.03 \\
        \hline
    \end{tabular}
    \caption{Mechanical properties of our best industrial-scale sample. The specimen Clax 2-3 a was subjected to cooling in LN$_2$, while the other one (Clax 2-3b) was not. Therefore, the exposure of our samples to very high thermal cycles does not influence the mechanical properties.}
    \label{tab:Clax YM}
\end{table}
To calculate the thermal contraction coefficient, a test was conducted, which consists in cooling a homogeneous sample of PMMA doped with \ce{Gd2O3} in LN$_2$, which is then left to thermalize for a few hours. Then the dimensions of this sample are measured with a caliper, after having quickly extracted it from the cryogenic bath. The dimensions are then compared to those that had been measured at room temperature, prior to cooling. This measurement was made on a parallelepiped taken from the Clax 2-3 industrial sample.
For each face of the parallelepiped we measured the width and length, along two axes perpendicular to each other (named "a" and "b") and parallel to the face under examination.
Then, the coefficient of linear thermal expansion (CLTE) was calculated according to Eq.~\ref{thermal_coefficient}
\begin{equation}\label{thermal_coefficient}
    \alpha_{L} = \frac{1}{L} \cdot \frac{dL}{dT}
\end{equation}
where L is the length of the sample considered and T is the temperature. 
The results are reported in Tab.~\ref{tab:shrinkage_test}.
\begin{table}[!h]
\centering
\begin{tabular}{ccccc}
\hline
\hline
\textbf{Face:}   & \textbf{Axis:} & \begin{tabular}[c]{@{}c@{}}\textbf{Length} ($\bm{T_{amb}}$)~\\{[}mm]:\end{tabular} & \begin{tabular}[c]{@{}c@{}}\textbf{Length} ($\bm{T_{LN_2}}$)~~\\{[}mm]:\end{tabular} & \begin{tabular}[c]{@{}c@{}}$\bm{\alpha_{L}}$\\{[}K$^{-1}$]:\end{tabular}  \\
\hline \hline
\multirow{2}{*}{Bottom} & a & 104.7 & 103.95    & (3.24 $\pm$ 0.06)$\cdot$10$^{-5}$    \\
     & b & 107.53     & 106.62    & (3.83 $\pm$ 0.06)$\cdot$10$^{-5}$    \\
\multirow{2}{*}{Top}    & a & 105.24     & 103.68    & (6.7 $\pm$ 0.07)$\cdot$10$^{-5}$     \\
     & b & 107.58     & 106.43    & (4.84 $\pm$ 0.07)$\cdot$10$^{-5}$    \\
1    & a & 117.86     & 116.24    & (6.22 $\pm$ 0.07)$\cdot$10$^{-5}$    \\
2    & a & 117.88     & 116.37    & (5.80 $\pm$ 0.07)$\cdot$10$^{-5}$    \\
3    & a & 117.81     & 116.64    & (4.49 $\pm$ 0.06)$\cdot$10$^{-5}$    \\
4    & a & 117.87     & 116.62    & (4.80 $\pm$ 0.06)$\cdot$10$^{-5}$    \\ \hline
\end{tabular}
\caption{Thermal contraction measurements on a specimen taken from the industrial sample Clax 2-3.}
\label{tab:shrinkage_test}
\end{table}
After that, the average value of all the measurements was calculated to obtain an estimate of the contraction coefficient, which is equal to (4.99 $\pm$ 0.02)$\cdot$10$^{-5}$~K$^{-1}$, in line with what is reported in the literature for the pure PMMA \cite{CTE_PMMA}.

\subsection{Industrial samples radiopurity}
In addition to the screening of the three main ingredients of \ce{Gd}-PMMA (MMA, \ce{Gd2O3} and Igepal CO-520\textsuperscript{\textregistered}) reported in Sec.~\ref{sec:Radio_ing}, we also measured a sample of Gd-PMMA from Clax s.r.l. to evaluate any eventual contamination that may be caused by the industrial production process. The results are reported in Tab.~\ref{tab:HPGe_Clax_PMMA}.
\begin{table}[!h]
\centering
\begin{tabular}{lc}
\hline
\hline
\textbf{Isotope}              & \begin{tabular}[c]{@{}c@{}}\textbf{A}\\ {[}mBq/kg{]}\end{tabular} \\
\hline \hline
$^{235 }$U                & $<$ 0.64                                           \\
$^{238}$U/$^{234m}$Pa       & $<$ 17                                             \\
$^{238}$U/$^{226}$Ra         & $<$ 0.26                                             \\
$^{232}$Th/$^{228}$Ac        & 0.4 $\pm$ 0.2                                           \\
$^{232}$Th/$^{228}$Th        & 0.4 $\pm$ 0.2                                        \\
$^{40}$K                 & 14 $\pm$ 3                                                 \\
$^{137}$Cs               & $<$ 0.24                                                                                 \\ \hline
\end{tabular}
\caption{Screening performed with HPGe detector on Clax gadolinium loaded PMMA. The sample that has been screened is "Clax 2-3", see Tab. \ref{tab:clax_samp} for the production details.}
\label{tab:HPGe_Clax_PMMA}
\end{table}
The sample under screening was produced during the second cycle of tests at this company, following the optimized procedure to obtain large mass thick sheets described before.\\
It is important to underline that the surfactant amount used to obtain this sample is 0.1$\%_{w}$ with respect to the MMA initial mass, but the surfactant had not been purified following the procedure described in Sec.~\ref{sec:K_reduction}. As anticipated the surfactant reduction alone already shows good results, but moreover we are reasonably sure that by applying the purification procedure the requirements reported in Tab.\ref{tab:GdProp} could be satisfied. This could be one of the future developments following this work.\\
In conclusion, considering the activities of the main ingredients and one of the final samples we can state that the production on the industrial line does not cause any significant contamination and is therefore usable for the production of Gd-PMMA for an experiment with stringent radiopurity requirements as DarkSide-20k.

%% file: Conclusions.tex
\section{Conclusions}
The goal of this work, carried on within the context of DarkSide-20k as one of the several R\&Ds, was to develop a new plastic hybrid material, made of a polymeric matrix rich in hydrogen and homogeneously loaded with gadolinium in high concentration. The material has to be ultra-pure from the radioactivity point of view and capable to survive at cryogenic temperatures. The strategy adopted was to use gadolinium oxide nanoparticles to create a mechanical dispersion in liquid MMA and then polymerize the mixture.\\ 
To minimize the clustering of the nanoparticles and the eventual sedimentation, the gadolinium oxide was functionalized with a commercial non-ionic surfactant, which introduces repulsive forces and steric hindrance. The efficiency of this first step of the procedure has been evaluated by performing FTIR and DLS measurements. The results indicate the presence of the surfactant on the nanoparticles and that the stability of the \ce{Gd2O3} dispersion is of the order of one hour.\\
The polymerization procedure, done in two steps, using two different chemical initiators, was optimized to obtain uniform thick samples, with a concentration of \ce{Gd} up to 2\%$_w$. \\
The hybrid material was tested to verify if the presence of nanoparticles influenced the thermo-mechanical properties of the polymeric matrix. DSC measurements show that the glass transition temperature does not change due to the presence of the treated nanoparticles. \\
The uniformity of gadolinium oxide distribution was verified for nearly all laboratory-scale samples, using the calcination technique. The results are satisfactory for sample heights up to 20 cm: this proves the feasibility of the process, to obtain 17 cm thick samples compliant with DarkSide-20k requirements.
Also, the mechanical characterization of the laboratory scale samples show that the presence of nanoparticles does not spoil the analyzed mechanical properties with respect to the pure PMMA. The samples have also passed the cryogenic tests. 
\\
The developed mixing procedure was successfully transferred to a partner company, Clax s.r.l., where it was optimized to obtain 12 cm thick sheets. These industrial samples have been characterized following the procedures developed during the laboratory phase, and have met all the requirements. 
In particular, the \ce{Gd2O3} uniformity, which was the most critical aspect, has been improved with respect to the laboratory samples and it is fully compatible with the requirements. 
Finally, we were able to reach the required levels of radiopurity, both for the single ingredients and for the final Gd-PMMA. 
In conclusion, the R\&D project led to the full development and characterization of this new radiopure neutron tagging material. The technology is  robust and based on the use of ingredients fully available on the market. The procedure, validated at the industrial level and suitable for the production of large-scale amount of material, has been patented in Italy (Patent for industrial invention n. 102021000028130, classification C08F) and is in the final stages of evaluation for patenting in the USA, EU and China \cite{Brevetto}.

%% file: Appendix.tex
\section{Nanoparticle distribution uniformity measurements}
Tab.~\ref{tab:calcinations} shows the results of the calcination characterization (reported in Sec.~\ref{sec:Calcinations}) on some laboratory samples. These samples were obtained using the same procedure (illustrated in Sec.~\ref{section:LabTests}), and differ in thickness (from $\approx$3 cm to $\approx$20 cm) and in the concentration of \ce{Gd2O3} nano-grains. For all, the homogeneity of the distribution of nano-grains along the vertical axis is satisfactory with respect to the requirements of Tab.~\ref{tab:GdProp}.
\begin{longtable}{ccccc}
\hline
\hline
\textbf{Sample} & \textbf{Section} & \textbf{\begin{tabular}[c]{@{}c@{}}Nominal \ce{Gd2O3} \\ concentration\\ \textnormal{{[}\%$_{w}${]}}\end{tabular}} & \textbf{\begin{tabular}[c]{@{}c@{}}Measured \ce{Gd2O3} \\ concentration\\ \textnormal{{[}\%$_{w}${]}}\end{tabular}} & \textbf{\begin{tabular}[c]{@{}c@{}}Surfactant \\ concentration\\ \textnormal{{[}\%$_{w}${]}}\end{tabular}} \\
\hhline{=====}
\endfirsthead
\multicolumn{4}{c}%
{{\bfseries Table \thetable\ continued from previous page}} \\
\endhead
\multirow{2}{*}{09-06-1} & Section 1 & 1 & 1.491 $\pm$ 0.003 & 1  \\
                         & Section 2 & 1 & 1.185 $\pm$ 0.003 & 1  \\ \hline 
\multirow{3}{*}{19-05-2} & Section 1 & 1 & 1.115 $\pm$ 0.003 & 1  \\
                         & Section 2 & 1 & 0.914 $\pm$ 0.003 & 1 \\
                         & Section 3 & 1 & 1.503 $\pm$ 0.003 & 1 \\ \hline 
\multirow{3}{*}{20-05-1} & Section 1 & 1 & 0.979 $\pm$ 0.003 & 1 \\
                         & Section 2 & 1 & 0.855 $\pm$ 0.003  & 1\\
                         & Section 3 & 1 & 1.288 $\pm$ 0.003  & 1\\ \hline 
\multirow{5}{*}{13-10-1} & Section 1 & 2.3 & 2.662 $\pm$ 0.003 & 2  \\
                         & Section 2 & 2.3 & 2.585 $\pm$ 0.004 & 2 \\
                         & Section 3 & 2.3 & 2.624 $\pm$ 0.003 & 2 \\
                         & Section 4 & 2.3 & 2.570 $\pm$ 0.003 & 2 \\
                         & Section 5 & 2.3 & 1.453 $\pm$ 0.003 & 2 \\ \hline 
\multirow{2}{*}{03-11-1} & Section 1 & 1 & 1.076 $\pm$ 0.003 & 0.1  \\
                         & Section 2 & 1 & 1.308 $\pm$ 0.003 & 0.1\\ \hline 
\multirow{4}{*}{05-10-1} & Section 1 & 2.3 & 3.007 $\pm$ 0.003 & 2   \\
                         & Section 2 & 2.3 & 2.649 $\pm$ 0.002 & 2   \\
                         & Section 3 & 2.3 & 2.423 $\pm$ 0.003 & 2   \\
                         & Section 4 & 2.3 & 2.501 $\pm$ 0.007 & 2   \\ \hline 
\multirow{4}{*}{29-09-1} & Section 1 & 1 & 1.278 $\pm$ 0.002 & 1  \\
                         & Section 2 & 1 & 1.246 $\pm$ 0.003 & 1  \\
                         & Section 3 & 1 & 1.243 $\pm$ 0.003 & 1 \\
                         & Section 4 & 1 & 1.165 $\pm$ 0.003 & 1 \\ \hline 
\multirow{4}{*}{26-05-1} & Section 1 & 1 & 1.070 $\pm$ 0.003 & 1 \\
                         & Section 2 & 1 & 0.995 $\pm$ 0.002 & 1 \\
                         & Section 3 & 1 & 1.014 $\pm$ 0.003 & 1 \\
                         & Section 4 & 1 & 1.146 $\pm$ 0.002 & 1 \\ \hline 
\multirow{4}{*}{09-12-2} & Section 1 & 1 & 1.104 $\pm$ 0.004 & 0.02 \\
                         & Section 2 & 1 & 0.999 $\pm$ 0.004 & 0.02 \\
                         & Section 3 & 1 & 1,036 $\pm$ 0.004 & 0.02 \\ 
                         & Section 4 & 1 & 1.029 $\pm$ 0.007 & 0.02 \\\hline 
\multirow{4}{*}{26-05-2} & Section 1 & 1 & 1.081 $\pm$ 0.002 & 1 \\
                         & Section 2 & 1 & 1.020 $\pm$ 0.004 & 1 \\
                         & Section 3 & 1 & 0.995 $\pm$ 0.004 & 1 \\ 
                         & Section 4 & 1 & 1.130 $\pm$ 0.002 & 1 \\ \hline 
\multirow{6}{*}{09-12-1} & Section 1 & 1 & 1.037 $\pm$ 0.002 & 0.1 \\
                         & Section 2 & 1 & 1.132 $\pm$ 0.002 & 0.1 \\
                         & Section 3 & 1 & 1.140 $\pm$ 0.002 & 0.1 \\
                         & Section 4 & 1 & 1.266 $\pm$ 0.002 & 0.1 \\
                         & Section 5 & 1 & 1.230 $\pm$ 0.003 & 0.1 \\
                         & Section 6 & 1 & 1.314 $\pm$ 0.003 & 0.1 \\ \hline 
\caption{Results of calcination tests carried out on some laboratory samples. It should be noted that the developed procedure allowed us to reach a satisfactory \ce{Gd2O3} homogeneity for different concentration values of the nano-grains and of the surfactant. Furthermore, the samples shown in the table also differ in thickness, in a range between 3 cm and 20 cm.}
\label{tab:calcinations}\\
\end{longtable}